\documentclass[pra,twocolumn,superscriptaddress,aps,showpacs,amsmath,amssymb,nofootinbib]{revtex4}
\usepackage{graphicx}
\usepackage{dcolumn}
\usepackage{bm}
\usepackage{color}

\newcommand{\bce}{\begin{center}}
\newcommand{\ece}{\end{center}}
\newcommand{\beq}{\begin{equation}}
\newcommand{\eeq}{\end{equation}}
\newcommand{\beqa}{\begin{eqnarray}}
\newcommand{\eeqa}{\end{eqnarray}}

\begin{document}

\title{Measure synchronization in quantum many-body systems}
\date{\today}

\author{Haibo Qiu}
\affiliation{College of Science, Xi'an University of Posts and
Telecommunications, 710121 Xi'an , China}
\affiliation{Departament d'Estructura i Constituents de la Mat\`{e}ria,\\
Universitat de Barcelona, 08028 Barcelona, Spain}

\author{Bruno Juli\'a-D\'\i az}
\affiliation{Departament d'Estructura i Constituents de la Mat\`{e}ria,\\
Universitat de Barcelona, 08028 Barcelona, Spain}

\author{Miguel Angel Garcia-March}
\affiliation{Departament d'Estructura i Constituents de la Mat\`{e}ria,\\
Universitat de Barcelona, 08028 Barcelona, Spain}

\author{Artur Polls}
\affiliation{Departament d'Estructura i Constituents de la Mat\`{e}ria,\\
Universitat de Barcelona, 08028 Barcelona, Spain}

\begin{abstract}

The concept of measure synchronization between two coupled quantum
many-body systems is presented. In general terms we consider two
quantum many-body systems whose dynamics gets coupled through the
contact particle-particle interaction. This coupling is shown to
produce measure synchronization, a generalization of synchrony to a
large class of systems which takes place in absence of dissipation.
We find that in quantum measure synchronization, the many-body
quantum properties for the two subsystems, e.g., condensed fractions
and particle fluctuations, behave in a coordinated way. To
illustrate the concept we consider a simple case of two species of
bosons occupying two distinct quantum states. Measure
synchronization can be readily explored with state-of-the-art
techniques in ultracold atomic gases and, if properly controlled, be
employed to build targeted quantum correlations in a sympathetic
way.
\end{abstract}


\maketitle

\section{Introduction}

Since its discovery by Huygens when observing coupled pendula in the
17th century~\cite{Pen}, synchronization has been described in
physics, chemistry, biology and even social
behavior~\cite{Review,Review2,Review4}, becoming a paradigm for
research of collective dynamics. It has been thoroughly studied in
classical nonlinear dynamical systems~\cite{Review3}, and extended
to chaotic ones~\cite{Chaos}. Only recently synchronization has been
studied in quantum systems, e.g., two coupled quantum harmonic
oscillators~\cite{Syn6}, a qubit coupled to a quantum dissipative
driven oscillator~\cite{Syn3}, two dissipative spins~\cite{Syn4},
and two coupled cavities~\cite{Syn5}. Last year, connections between
quantum entanglement and synchronization have been discussed in
continuous variable systems ~\cite{Zam2, Syn7}.

A decade ago, Hampton and Zanette introduced a new concept termed
measure synchronization (MS) for coupled Hamiltonian
systems~\cite{MS99}. They found that two coupled Hamiltonian systems
experience a synchronization transition from a state in which the
two subsystems visit different phase space regions to a state in
which ``their orbits cover the same region of the phase space with
identical invariant measures''~\cite{MS99}. The control parameter is
the coupling strength between the two subsystems.

The key difference between MS and conventional synchronization is
that MS takes place in absence of dissipation. In standard
synchronization dissipation plays a key role, as it is responsible
for the collapse of any trajectory of the system in phase space. For
coupled Hamiltonian systems, phase space volume must be conserved
following Liouville's theorem, thus preventing the collapse of any
trajectory in phase space. In the case of MS, two coupled
Hamiltonian systems become synchronized when they cover the same
phase space domain, without requiring that the synchronized systems
have the same evolution trajectories.

In this article we introduce measure synchronization, a concept up
to now only considered in a classical framework, into the quantum
many-body regime. First exploratory studies of measure
synchronization in quantum systems have been done in different
contexts, i.e., MS has been discussed in coupled Hamiltonian systems
associated with nonlinear Schr\"odinger equations~\cite{MS05}; also,
MS transitions have been revealed on meanfield theories describing
condensed bosonic quantum many-body
systems~\cite{MS03,Qiu10,MS12,Qiu13}. However, it is worth stressing
that in the above cases the dynamical variables describing these
quantum systems are classical, i.e. quantum fluctuations are
neglected, a reasonable approximation in bosonic systems which are
fully condensed~\cite{Leggett01}. This made the correspondence
between the classical MS concept introduced in Ref.~\cite{MS99} and
the MS studies of these quantum systems straightforward. So
conceptually, measure synchronization discussed in these contexts
remained classical. Here, we tackle the problem in a fully many-body
quantum mechanical way. A major conceptual difference is that, in
the general case we need to identify quantum many-body observables
which allow us to characterize MS-like behaviors provided the very
definition of the area covered by each subsystem in phase space is
absent.

We consider two quantum many-body systems (QMBS) which are coupled
through a local interaction term. Our main finding is that we
characterize a crossover behavior from non-MS to MS in the evolution
of the quantum many-body properties of the subsystems. This implies
that two QMBS, which if non coupled would develop different quantum
correlations, will, if sufficiently coupled, have similar condensed
fractions, particle fluctuations, etc. This is an effect which will
affect the behavior of future QMBS and quantum simulators, and
which, if properly controlled, can be employed to share or to induce
quantum correlations between different degrees of freedom in the
system. MS is a dynamical feature which we will show to appear in
the evolution of QMBS. It describes how, under certain premises, the
dynamics of two weakly coupled quantum subsystems becomes coherent
after a short transient time. MS describes how two subsets of a QMBS
will evolve in a collective way, exchanging energy during the full
evolution, exploring similar average values of relevant observables
and developing similar quantum correlations.

It is worth emphasizing that our ability to understand and utterly
control quantum correlations in QMBS is the key to producing
powerful technological applications. A notable recent example is the
case of pseudo-spin squeezed states~\cite{WI94}, which can be
produced in bosonic Josephson junctions~\cite{esteve08}. In this
case, producing fragmented ultracold gases is shown to notably raise
the precision achievable in quantum metrology
experiments~\cite{gross10,riedel10}. These applications will become
a reality in the near future thanks to the miniaturization of
ultracold atomic systems~\cite{nshii13}. As shown here, MS can be
used to transfer, or sympathetically produce, fragmentation in one
subsytem of the QMBS which can, for instance, improve
interferometric signals.

The article is organized in the following way. In Sec.~\ref{sec:teo}
we describe the many-body Hamiltonian. In Sec.~\ref{sec:res} we
present our results, concerning the onset of MS and how it shows in
the many-body properties of the system. In Sec.~\ref{sec:exp} we
sketch an experimental implementation with ultracold atomic gases. A
summary and conclusions are provided in Sec.~\ref{sec:sum}.

\section{Many-body hamiltonian}
\label{sec:teo}
 To illustrate the many-body quantum MS we consider the
simplest implementation we can think of. These are two different
kinds of bosons, $A$ and $B$, populating solely two quantum states,
$L$ and $R$. We will consider a linear coupling between the two
quantum states and contact interaction for $AA$, $AB$, and $BB$
bosons. The many-body Hamiltonian considered for $N_A$ and $N_B$
atoms in two modes is,
\begin{eqnarray}
\hat{H} &=& \hat{H}_A + \hat{H}_B + \hat{H}_{AB}, \label{eq:H}
\end{eqnarray}
where
\begin{eqnarray}
\hat{H}_A &=& \frac{U_{A}}{2}
\left[(\hat{a}_{L}^{\dag}\hat{a}_{L})^{2}
+(\hat{a}_{R}^{\dag}\hat{a}_{R})^{2}\right] -J_{A}(\hat{a}_{L}^{\dag
}\hat{a}_{R}+\hat{a}_{R}^{\dag }\hat{a}_{L})
\nonumber \\
\hat{H}_B&=&\frac{U_{B}}{2}
\left[(\hat{b}_{L}^{\dag}\hat{b}_{L})^{2}+(\hat{b}_{R}^{\dag}\hat{b}_{R})^{2}\right]
-J_{B}(\hat{b}_{L}^{\dag }\hat{b}_{R}+\hat{b}_{R}^{\dag
}\hat{b}_{L})
\nonumber\\
\hat{H}_{AB} & =& U_{AB}
(\hat{a}_{L}^{\dag}\hat{a}_{L}\hat{b}_{L}^{\dag }\hat{b}_{L}
+\hat{a}_{R}^{\dag}\hat{a}_{R}\hat{b}_{R}^{\dag }\hat{b}_{R}) \,.
\end{eqnarray}
$\hat{a}_{L(R)}^{\dag }$ $(\hat{a}_{L(R)})$ and
$\hat{b}_{L(R)}^{\dag }$ $(\hat{b}_{L(R)})$ are creation
(annihilation) operators for the single-particle modes $L$ or $R$ of
the two species. The terms proportional to $J_{A(B)}$ are the linear
coupling terms, which in absence of any interaction would induce
periodic Rabi oscillations of the populations between the states $L$
and $R$. $U_A$, $U_B$, and $U_{AB}$ measure the $AA$, $BB$, and $AB$
contact interactions. The $U_{AB}$ term is the only one coupling the
dynamics of the $A$ and $B$ subsystems, and will be responsible for
the MS between both of them.

The Hamiltonian can be numerically diagonalized in the
$N_D=(N_{A}+1)(N_{B}+1)$ dimensional space spanned by the many-body
Fock basis tensor product of the $A$ and $B$ Fock states, e.g., for
the $A$, $|N_{A,L}\rangle\equiv 1/\sqrt{ N_{A,L}! N_{A,R}!} \
(\hat{a}^\dagger_L)^{N_{A,L}} (\hat{a}^\dagger_R)^{N_{A,R}} |{\rm
vac}\rangle$, with $N_{A,L}=0,\dots, N_{A}$ and
$N_{A,R}=N_{A}-N_{A,L}$. The most general $N$-particle state can be
written as
\begin{equation}
|\Psi\rangle= \sum_{N_{A,L}=0}^{N_A} \sum_{N_{B,L}=0}^{N_B}
c_{N_{A,L},N_{B,L}} \ |N_{A,L}, N_{B,L}\rangle\,.
\end{equation}
The time evolution of any given initial state is governed by the
time dependent Schr\"odinger equation, $i \, \hbar \,
\partial_t |\Psi(t)\rangle = \hat{H} |\Psi(t)\rangle$. Once we have
computed the many-body state, we can obtain average particle numbers
on modes $L$ and $R$, $ \langle N_{A,\alpha}\rangle= \langle \Psi|
\hat{a}^\dagger_{\alpha} \hat{a}_{\alpha} |\Psi\rangle $, $\langle
N_{B,\alpha}\rangle= \langle \Psi| \hat{b}^\dagger_{\alpha}
\hat{b}_{\alpha} |\Psi\rangle$, with $\alpha=L, R$. The imbalance of
population for each species is defined as
$Z_{A(B)}=(N_{A(B),L}-N_{A(B),R})/N_{A(B)}$.

To characterize the degree of condensation of each subsystem, $A$
and $B$, at any given time we will make use of the one-body density
matrix, $\rho$~\cite{ours10}. For a state $|\Psi\rangle$ it is
defined as, e.g., for species $A$, $ \rho^{A}_{ij}= \langle \Psi |
\hat{\rho}^{A}_{ij}| \Psi \rangle$, with $\hat{\rho}^{A}_{ij}=
a^\dagger_i a_j$, and $i,j=L,R$. The traces of $\rho^{A}$ and
$\rho^{B}$ are normalized to the number of atoms in each subsystem,
$N_{A}$ and $N_{B}$. The two normalized eigenvalues (divided by the
total number of atoms $N_{A}$) are $n_{a_1(a_2)}$, with $n_{a_1}\geq
n_{a_2}\geq 0$. We always have $n_{a_1}+n_{a_2}=1$. The larger
eigenvalue is also called the condensed fraction. Similar
definitions are used for species $B$.

\begin{figure}[t]
\includegraphics[width=1.3\columnwidth, angle=270]{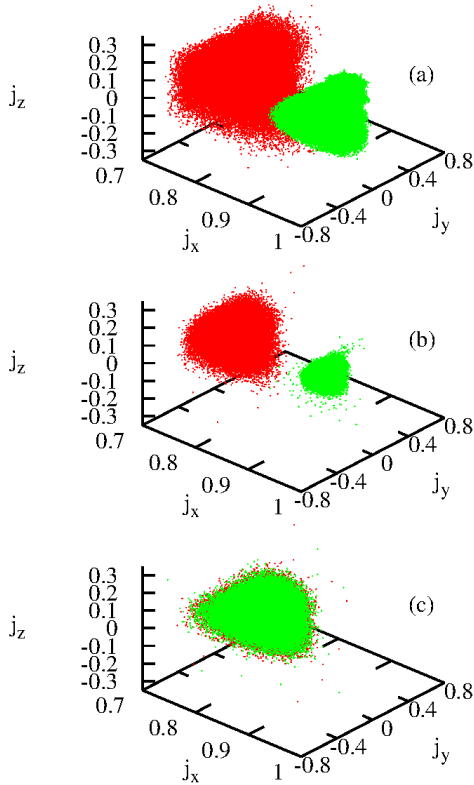}
\caption[]{(Color online) {\bf Measure synchronization.} Quantum
many-body measure synchronization characterized by the domain
covered during the evolution of each subsystem on the 3D space
defined by the average value of the pseudo-angular momentum,
$j_x\equiv (2/N) \langle \hat{J}_x\rangle$, $j_y\equiv (2/N) \langle
\hat{J}_y\rangle$, and $j_z\equiv (2/N) \langle \hat{J}_z\rangle$
for species $A$ (red) and $B$ (green). (a) $U_{AB}=0$ (non-MS), (b)
$U_{AB}=0.008\ U$ (non-MS), and (c) $U_{AB}=0.5\ U$ (MS). In panel
(c) the two clouds overlap. $Z_A(0)=0.2$, $Z_B(0)=0.4$, and
$N_A=N_B=30$. In all panels the points correspond to periodic
moments within one time evolution, we show them as dots, instead of
a continuous line, to avoid one of the species hiding the other one.
} \label{fig1}
\end{figure}

A way to characterize the transition from non-MS to MS dynamics in
classical systems is
by looking at the time average of the energies of subsystems $A$ and
$B$~\cite{Qiu13,MS03}. In MS dynamics, both subsystems cover, with
equal density, the same phase space domain, which reflects on equal
long-time averages of the energies of the subsystems defined as
\begin{equation}
\label{fdspect4} \bar{E}_{A(B)}  = {1\over T} \int_{0}^{T} \
E_{A(B)}(t) \ dt \ ,
\end{equation}
where the expectation values of the energy for each subsystem $A$
(and $B$) at time $t$ are $E_{A}(t)=\langle
\Psi(t)|\hat{H}_{A}|\Psi(t)\rangle $, and $E_{B}(t)=\langle
\Psi(t)|\hat{H}_{B}|\Psi(t)\rangle $, with $|\Psi(t)\rangle $ the
evolved quantum state.

\section{Results}
\label{sec:res}

We set both intraspecies interactions to be the same, i.e.~, $U
\equiv U_{A}=U_{B}$, with $N U/J=7.2$, and also choose
equal linear couplings, $J \equiv J_{A}=J_{B}$. 
We take as a unit of time the Rabi time, $t_{\rm Rabi}=\pi/J$, and
as a unit of energy, $\hbar/t_{\rm Rabi}$. Our initial states will
in all cases be coherent states for both the $A$ and $B$ species, in
which all atoms populate the single particle state
$(1/\sqrt{2})(\cos \theta/2 \ \hat{a}^\dagger_L + \sin \theta/2\
a^\dagger_R)$ [with initial population imbalance, $Z(0)=\cos
\theta$]. These states will evolve under the action of the many-body
Hamiltonian. We will look for a transition from non-MS to MS in the
collective dynamics of the many-body state as we vary the
interspecies interaction strength $U_{AB}$.

The transition from non-MS to MS dynamics is shown in
Fig.~\ref{fig1}. We plot the average value of the pseudo-angular
momentum operators which can readily be constructed from the
creation and annihilation operators of each
species~\cite{Leggett01}, $\hat{J}_x =(1/2)(\hat{a}^\dagger_L
\hat{a}_R+\hat{a}_L \hat{a}^\dagger_R)$, $\hat{J}_y =1/(2
i)(\hat{a}^\dagger_L \hat{a}_R-\hat{a}_L \hat{a}^\dagger_R)$,
$\hat{J}_z =(1/2)(\hat{a}^\dagger_L \hat{a}_L-\hat{a}^\dagger_R
\hat{a}_R)$. In our conditions, fixed $N_A$ and $N_B$, these
operators build the symmetric representation of $SU(2)$ of dimension
$N_{A}+1$ and $N_{B}+1$. As shown in the three-dimensional (3D)
figure, in the non-MS cases, $U_{AB}=0$ and $U_{AB}=0.008 \ U$, the
domains of ($\langle J_x\rangle$, $\langle J_y\rangle$, $\langle
J_z\rangle$) explored by each subsystem are disjointed. In the MS
case, however, both domains completely overlap. This feature can be
regarded as the many-body counterpart of the classical definition of
MS, in which the phase space domain covered by both subsystems is
the same.

\begin{figure}
\includegraphics[width=0.99\columnwidth, angle=0, clip=true]{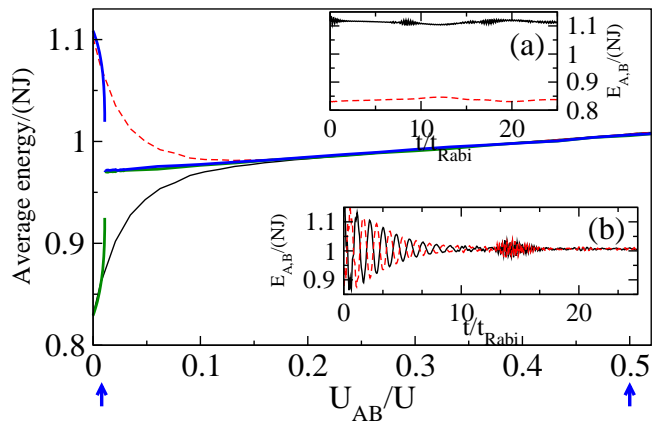}
\caption[]{(Color online) {\bf From non-MS to MS.} Long-time
averaged energies, Eq.~(\ref{fdspect4}), for the two species
$\bar{E}_A$ (solid) and $\bar{E}_{B}$ (dashed) as a function of
$U_{AB}$. MS dynamics corresponds to equal averages,
$\bar{E}_{A}=\bar{E}_{B}$. $Z_A(0)=0.2$, $Z_B(0)=0.4$, and
$N_A=N_B=30$. In the insets we depict $E_A(t)$ and $E_B(t)$ for two
specific values of $U_{AB}/U=0.008$ (a) and $U_{AB}/U=0.5$ (b). The
two values are marked with arrows in the main figure. $T=1000 \
t_{\rm Rabi}$. The green and blue lines correspond to the classical
prediction of Ref.~\cite{Qiu13}. } \label{fig2}
\end{figure}

In classical systems MS implies that both subsystems have similar
long-time averages of their energies. Importantly, the non-MS to MS
transition in classical systems is discontinuous, which allows one
to define a critical point to characterize the dynamical phase
transition~\cite{MS99}. This is seen in Fig.~\ref{fig2}, where we
depict the average energy $\langle E_{A}\rangle$ and $\langle
E_{B}\rangle$ as a function of the interspecies interaction
$U_{AB}$, with $\bar{E}_A=\bar{E}_B$ characterizing MS. The
classical results~\cite{Qiu13}, depicted in green and blue, feature
the known discontinuity. In the many-body case the situation is
different; the dynamical phase transition is replaced by a crossover
behavior, therefore no critical point can be unambiguously defined.
There is no criticality which involves logarithmic singularity in
the quantum measure synchronization as compared with classical
theory of measure synchronization~\cite{MS03,Qiu13}. Also note that
in the many-body case, MS appears at higher values of $U_{AB}$ as
compared to the classical transition.  The inset in Fig.~\ref{fig2}
shows the behavior of $E_A(t)$ and $E_B(t)$ for the two different
regions. In the MS case, the two subsystems exchange energy in such
a way that their energies oscillate around the same average value
maintaining an almost constant sum. In the non-MS dynamics, the
energies of the subsystems are never fully exchanged, and $A$ has
always more average energy than $B$. For different initial
conditions, particle numbers, and parameters $NU/J$ we obtain a
similar picture (see Appendix~\ref{app}), the main difference being
the size of the MS and non-MS regions.

\begin{figure}[t]
\includegraphics[width=1.\columnwidth, angle=0,clip=true]{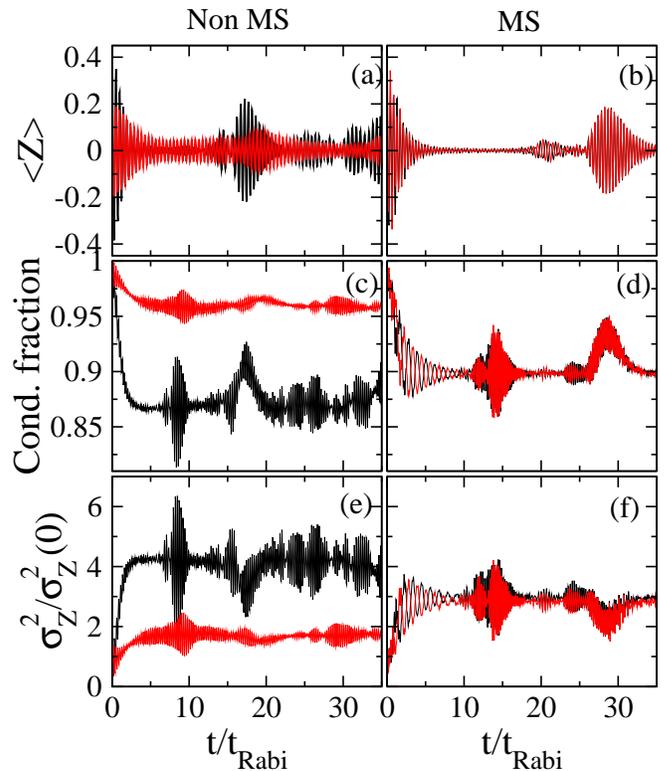}
\caption[]{(Color online) {\bf Measure synchronization on the
many-body properties.} We compare the properties of both subsystems
$A$ (black) and $B$ (red) as a function of time, for a non-MS
dynamics, $U_{AB}=0.008\ U$ (left panels) and for MS dynamics,
$U_{AB}=0.5\ U$ (right panels). The population imbalances [(a),
(b)], condensed fractions [(c), (d)], and dispersion of the
population imbalance $\sigma^2_Z=\langle Z^2\rangle -\langle
Z\rangle^2$ [(e), (f)]. All magnitudes show a signature of the
difference between the non-MS and MS dynamics. $Z_A(0)=0.2$,
$Z_B(0)=0.4$, and $N_A=N_B=30$.} \label{fig3}
\end{figure}

Now we concentrate on the evolution of the many-body properties of
both subsystems. In Fig.~\ref{fig3} we consider the same initial
state and two values of $U_{AB}$, $0.008\ U$ and $0.5\ U$, giving
rise to non-MS and MS, respectively. Figure~\ref{fig3}, panels (a)
and (b) show the population imbalance between the two quantum states
($L$ and $R$) for each subsystem. The quantum many-body evolution
becomes apparent, with characteristic collapse and revival
dynamics~\cite{ima97}. This collapse and revival dynamics has been
addressed for single component junctions~\cite{Milburn97, JAME06}
and it can be understood in finite systems due to the finite number
of frequencies entering in any dynamical evolution in the system.

We note that before reaching MS, the dynamics of the two subsystems
is different both in the amplitude of the oscillation and on the
times for collapses and revivals. After reaching MS, the times for
collapses and revivals are the same. This striking feature provides
a way to characterize quantum measure synchronization in the rhythms
of the coupled Hamiltonian systems. Furthermore, we note that the
oscillating amplitude for the two subsystems will be the same once
MS is achieved, which is a feature that also shows in classical
measure synchronization states~\cite{MS12}.

A crucial feature of quantum many-body bosonic systems is the
appearance of correlations stronger than those present in
Bose-Einstein condensed clouds. An initially condensed system loses
condensation during the evolution, and becomes
fragmented~\cite{baym} [see panels (c) and (d). This fragmentation
also takes place if there is no coupling between the
subsystems~\cite{ours10}. Interestingly, in the MS dynamics, the
condensed fraction of both subsystems gets clearly correlated after
a very short transient time, having the same envelope of the
oscillation amplitudes, which is a key feature of MS. This feature
is found with all particle numbers studied $N_A=N_B\leq 80$ (see
Appendix~\ref{app}). This similar behavior is also exhibited in the
dispersions of particle differences [Fig.~\ref{fig3}, panels (e) and
(f)]. This is of special significance, as this is directly related
to the emergence of cat-like many-body states or pseudo-spin
squeezed states in the evolution~\cite{ours10}. The latter provide a
direct application of this physics to improve our precision
measurements~\cite{esteve08}.

\section{Proposal for an experimental implementation}
\label{sec:exp}

The aforementioned MS can be studied with state-of-the-art
experimental techniques in ultracold atomic physics. The almost
perfect decoupling of ultracold atoms from their environment enables
the investigation of the quantum measure synchronization in
conservative systems. We describe a feasible system which can
simulate with good precision the many-body Hamiltonian in
Eq.~(\ref{eq:H}) using trapped ultracold atomic
gases~\cite{bloch-rmp,lewenstein-book}. We consider a two-species
ultracold atomic cloud trapped in a symmetric double-well potential.
In the weakly interacting regime, assuming the atom-atom
interactions are correctly described by a contact interaction, and
following similar steps as in Ref.~\cite{Milburn97}, one obtains
Eq.~(\ref{eq:H}). The classical predictions of Eq.~(\ref{eq:H}) have
been studied in
Refs.~\cite{Ashhab02,diaz09,satija09,mazzarella10,naddeo10,mmm11}
and some many-body features in Ref.~\cite{weiss}. The linear
couplings $J$ are proportional to the energy splitting between the
quasidegenerate ground state of the double-well potential. The
atom-atom interaction terms, are given by $U_\sigma=(4\pi\hbar^{2}
a_\sigma  /m_\sigma ) \int{\left|{\varphi_\sigma}\right|}^4 d{\bf
r}$, $U_{AB}= 2\pi\hbar^{2} a_{AB} (\frac{1}{{m_A}}+
\frac{1}{{m_{B}}})\int {\left|{\varphi _A} \right|}^2
\left|{\varphi_{B}} \right|^2 d{\bf r}$, where $\sigma$ refers to
atoms $A$ or $B$. $\varphi$ are localized single particle states;
the localized single particle states have been normalized as $\int
d{\bf r} |\varphi({\bf r})|^2 =1$. $a_\sigma$ is $s$-wave scattering
length between atoms $\sigma$, with mass $m_\sigma$. $a_{AB}$ is the
interspecies $s$-wave scattering length. The scattering lengths are
varied routinely in ultracold atom experiments by means of Feshbach
resonances~\cite{Fesh} or confinement induced
resonances~\cite{conf}. A possible specific experimental
implementation could be an external double-well potential as in
Ref.~\cite{Albiez05} or the double-well inside the quantum chip used
in Ref.~\cite{berrada13}. It has been shown that two mode BECs can
be prepared in coherent states experimentally~\cite{Zibold10,luc11}.
It would be possible to characterize quantum MS by investigating the
times of collapses and revivals for the two species. There are also
other experimental options for consideration, i.e., microcavity
exciton-polaritons system~\cite{car13}, or coupled micropillars
system~\cite{gal13}. Even though ultracold atomic samples are well
isolated from the environment, there is one source of decoherence
which could affect the onset of MS to non-MS transitions. This is
the presence of losses in the system. These have been studied in
detail for single component Josephson junctions, finding a
constraint on the maximum attainable correlation which can be
produced in the junction~\cite{li08}. A study of the effect of
losses on the MS to non-MS transition falls beyond the scope of the
present article.

\section{Summary and conclusions}
\label{sec:sum}

We introduced the concept of measure synchronization in quantum
many-body systems. To exemplify the phenomenon we have considered a
two-species bosonic Josephson junction made of a small number of
atoms which can be experimentally studied in a number of different
setups. Importantly, the measure synchronization occurs at the
many-body quantum level, showing how properties such as the
condensed fraction or the fluctuations in particle number of the two
species behave coordinately above a certain coupling strength
between the two systems. The findings reported apply
to a variety of quantum many-body systems. An important application
which can be envisaged is to profit from the MS described here
to build targeted quantum correlations of certain
degrees of freedom in the system in a sympathetic way. That is, when
one can experimentally control and prepare the quantum correlations
in one part of the system (e.g. one of the species), this method can
be used to build similar quantum correlations in other parts of the
system which cannot be experimentally controlled (e.g. the
other species). In this MS regime, different parts of the system
will develop similar quantum correlations and other quantum
properties after a short transient time.

\begin{acknowledgments}
The authors thank J. Martorell for useful comments on the
manuscript. This work was supported by the National Natural Science
Foundation of China (Grant No. 11104217). We also acknowledge
partial financial support from the DGI (Spain) (Grant
No.FIS2011-24154) and the Generalitat de Catalunya (Grant No.
2009SGR-1289). B.J.-D. is supported by the Ram\'on y Cajal program.
\end{acknowledgments}

\appendix

\section{Classical and full quantum descriptions}
\label{app}

\begin{figure}[t]
\includegraphics[width=0.99\columnwidth, angle=0, clip=true]{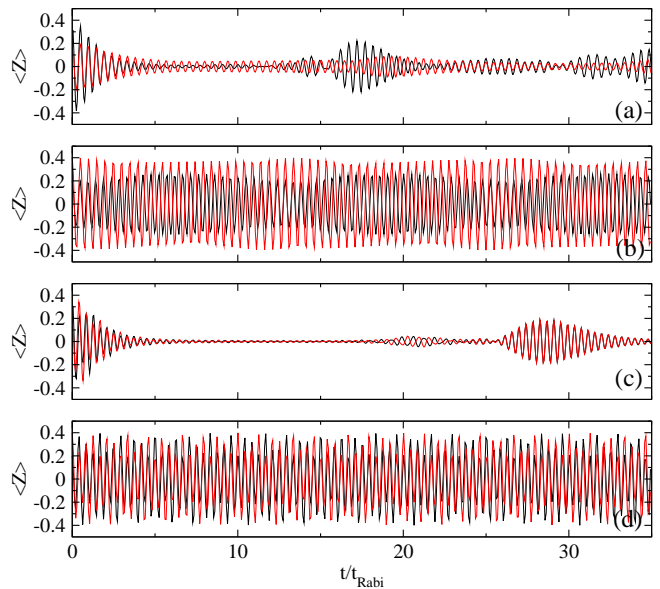}
\caption[]{(Color online) Evolution of the population imbalance of
both species. Comparison between the full quantum results shown in
Fig.~3 and the classical predictions of Ref.~\cite{Qiu13}. (a) and
(b) are for the full quantum and classical prediction for a non-MS
case ($U_{AB}=0.008 U$), respectively. (c) and (d) are the full
quantum and classical prediction for the MS case ($U_{AB}=0.5 U$),
respectively. \label{figa1}}
\end{figure}

In this Appendix we analyze, by considering increasingly larger
particle numbers, the relation between the classical and full
quantum descriptions. The main interest of the present article is to
extend the concept of measure synchronization to systems which do
not accept a full classical description. Thus, we have emphasized
the effects on the magnitudes which have no direct classical analog,
such as the condensed fractions and the fluctuations of the particle
numbers, shown in Figs.~\ref{fig3}(c) and ~\ref{fig3}(d) and Figs.
~\ref{fig3}(e) and ~\ref{fig3}(f), respectively. Interestingly,
particle number fluctuations can be measured experimentally and
provide a good way of pinning down correlated states in these
systems~\cite{esteve08,Zibold10}.

To take the classical limit in a meaningful way we will perform
exact numerical simulations for $N_A=N_B\leq 80$, keeping both $J$
and $NU/J=7.2$ constant, and compare to the corresponding classical
predictions. As occurred in the case of a usual bosonic Josephson
junction, the most remarkable difference between the classical and
quantal results is the presence of quantum collapses and revivals in
the latter~\cite{Milburn97}. This can be seen already on the
evolution of the average values of the particle number imbalances of
species A and B. In Fig.~\ref{figa1} we depict the comparison
between the average particle imbalance of each species reported in
Fig.~3 (obtained for $N=30$) and the corresponding classical
prediction~\cite{Qiu13}. Also, this is one of the signatures that
shows that MS can be characterized by the rhythms observed in the
dynamical evolution of the two coupled subsystems. In contrast,
classical measure synchronization is characterized by a spatial
localization in the phase space associated to the conjugate
variables describing the imbalance and the phase difference of each
subsystem with no need of synchronization in the time evolution of
the variables of each subsystem~\cite{Qiu10}.

\begin{figure}[t]
\includegraphics[width=0.99\columnwidth, angle=0, clip=true]{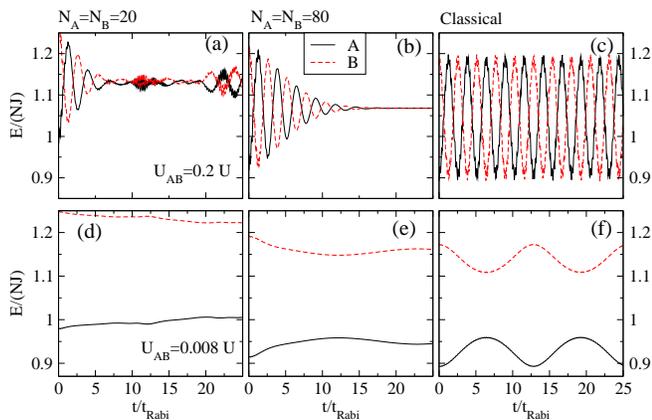}
\caption[]{(Color online) Evolution of the energy of each subsystem.
The upper and lower rows correspond to a MS and a non-MS situation,
respectively. The value of $U_{AB}=0.008$ U for non-MS, and
$U_{AB}=0.2$ U for the MS case. Exact many-body results for
$N_A=N_B=20$ and $N_A=N_B=80$ are given panels [(a), (d)] and [(b),
(e)], respectively. Panels [(c), (f)] contain the classical
predictions~\cite{Qiu13}. $Z_A(0)=0.2$ and $Z_B(0)=0.4$.}
\label{figa2}
\end{figure}

\begin{figure}[t]
\includegraphics[width=0.99\columnwidth, angle=0, clip=true]{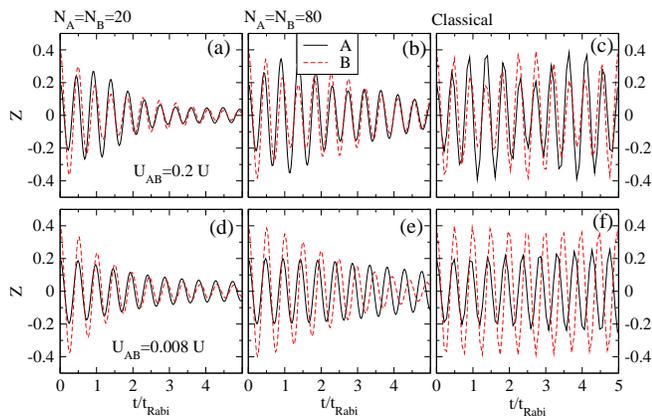}
\caption[]{(Color online) Evolution of the population imbalance for
both species. The upper and lower rows correspond to a MS and a
non-MS situation, respectively. The value of $U_{AB}=0.008$ U for
non-MS, and $U_{AB}=0.2$ U for the MS case. Exact many-body results
for $N_A=N_B=20$ and $N_A=N_B=80$ are given panels [(a), (d)] and
[(b), (e)], respectively. Panels [(c), (f)] contain the classical
predictions~\cite{Qiu13}. $Z_A(0)=0.2$ and $Z_B(0)=0.4$.}
\label{figa2a}
\end{figure}

As expected, increasing the number of particles, the classical
predictions better describe the initial behavior of the quantum
ones. In Fig.~\ref{figa2} we present the average values of the
energy of each subsystem as a function of time comparing the
classical results to quantum ones at $N_A=N_B=20$ and 80. In the MS
case $(U_{AB}=0.2 U)$, panels (a)--(c), the classical result shows
quasiperiodic oscillations for both $E_A(t)$ and $E_B(t)$. As
expected for the MS case, both subsystems have the same mean energy,
when averaged over long times. This feature is also present in the
quantum calculation, as shown in Fig.~2 for $N_A=N_B=30$, already
for $N_A=N_B=20$, Fig.~\ref{figa2} (a). In this case the oscillation
is clearly damped for times around $5 t_{\rm Rabi}$, departing from
the classical results fairly early. As the total number of particles
is increased to
 $N_A=N_B=80$, the time of the first collapse increases
$\simeq 10 t_{\rm Rabi}$. In Fig.~\ref{figa2a} we depict the evolution
of the average population imbalance in both situations, which clearly
shows how the classical result improves the description of the initial
time evolution as $N$ is increased.

In the non-MS situation the classical result departs earlier from
the quantal predictions [see Fig.~\ref{figa2}(d)--\ref{figa2}(f)].
In this case, the classical case clearly shows different
long-time-averaged values of the energies for each subsystem, a
feature of non-MS. In the quantum results this is also observed,
albeit in this case even for $N_A=N_B=80$ the classical and quantum
results differ quantitatively already for times of the order of $3
t_{\rm Rabi}$. Note the collapses and revivals inherent to the
quantum description make it difficult to talk about long time
averages of the signals.

\begin{figure*}[t]
\includegraphics[width=1.8\columnwidth, angle=0, clip=true]{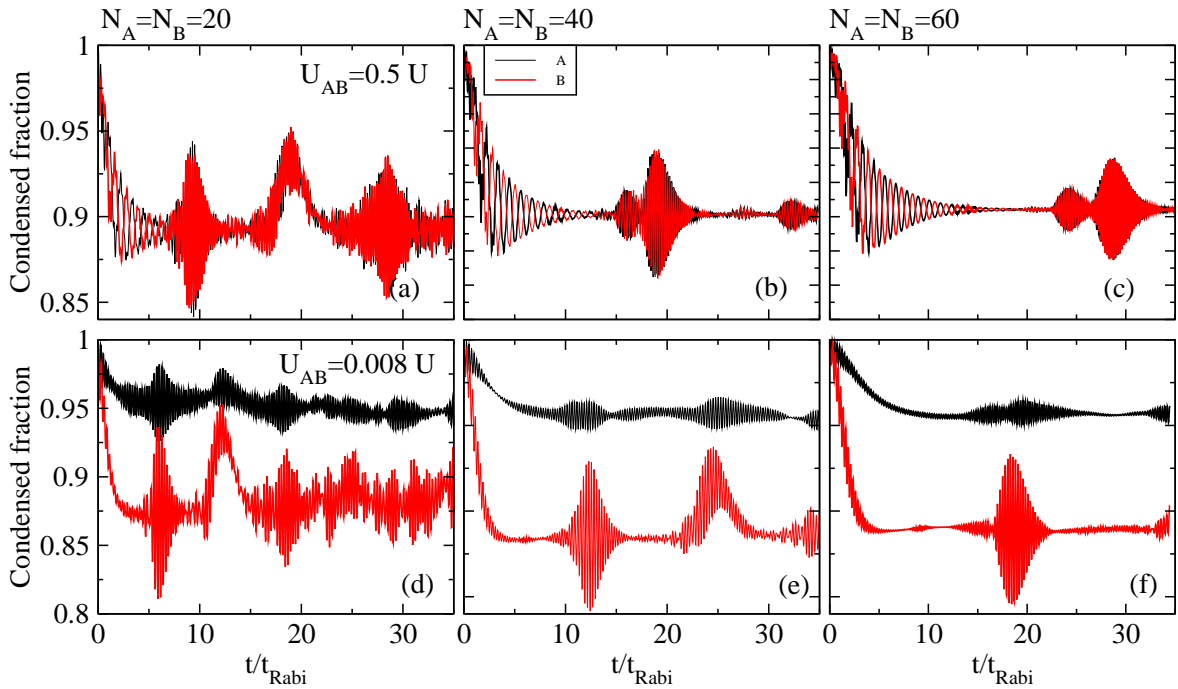}
\caption[]{(Color online) The upper and lower rows correspond to a
MS and a non-MS situation, respectively. The value of $U_{AB}=0.008
U$ is for non-MS, and $U_{AB}=0.5 U$ is for the MS case. Exact
many-body results for $N_A=N_B=20$, 40 and 60, are given in panels
[(a),(d)], [(b),(e)] and [(c),(f)], respectively. $Z_A(0)=0.2$ and
$Z_B(0)=0.4$.} \label{figa3}
\end{figure*}

As discussed above, we find a synchronization of the fragmentation
of the subsystems in the MS case as opposed to the non-MS
situations. In the classical description this is of course absent,
as the subsystems are fully condensed during the evolution. In the
quantum case even for small number of particles $N_A=N_B=20$ we find
this clear feature [see Fig.~\ref{figa3}(a) and
Fig.~\ref{figa3}(d)]. In the MS case both subsystems clearly
fragment in a synchronized way [Figs.~\ref{figa3}(a)
-~\ref{figa3}(c)] as opposed to the non-MS case
[Figs.~\ref{figa3}(d)--\ref{figa3}(f)]. Note also that MS produces
more overall fragmentation in the system, as it is the less
fragmented component A, the one that follows the more fragmented one
B. The time scale in which the system fragments is found to be
mostly independent of the number of particles, for the particle
numbers considered. At $t\simeq 10 t_{\rm Rabi}$ the maximum degree
of correlation is already built in the system. Also the amount of
fragmentation is found to be almost independent of the particle
number considered, although as expected it decreases slowly with
particle number.


\begin{thebibliography}{9}

\bibitem{Pen} C. Huygens, \textit{Horologium Oscillatorium} (Apub F. Muquet, Parisiis, 1673).

\bibitem{Review} Y. Kuramoto, \textit{Chemical Oscillations, Waves and Turbulence} (Springer,
Berlin) (1984).

\bibitem{Review2} Z. N\'eda, E. Ravasz, Y. Brechet, T. Vicsek, and A.-L. Barab\'asi,  Nature (London),  {\bf 403}, 850 (2000).

\bibitem{Review4}
A. Arenas, A. D\'iaz-Guilera, J Kurths, Y. Moreno, and C. Zhou,
Phys. Rep. {\bf 469} (3), 93 (2008).

\bibitem{Review3} A. Pikovsky, H. Rosenblum and J. Kurths,\textit{
Synchronization. A Universal Concept in Nonlinear Sciences}
(Cambridge University Press, Cambridge, England, 2001).

\bibitem{Chaos} L. M. Pecora and T. L. Carroll,
Phys. Rev. Lett. {\bf 64}, 821 (1990).

\bibitem{Syn6}
G. L. Giorgi, F. Galve, G. Manzano, P. Colet, and R. Zambrini,
Phys. Rev. A {\bf 85}, 052101 (2012).

\bibitem{Syn3}
O. V. Zhirov and D. L. Shepelyansky,
Phys. Rev. B {\bf 80}, 014519 (2009).

\bibitem{Syn4}
P. P. Orth, D. Roosen, W. Hofstetter, and K. LeHur,
Phys. Rev. B {\bf 82}, 144423 (2010).

\bibitem{Syn5}
T. E. Lee and M. C. Cross,
Phys. Rev. A {\bf 88}, 013834 (2013).

\bibitem{Zam2}  G. Manzano, F. Galve, G. L. Giorgi,
E. Hern\'andez-Garc\'ia, and R. Zambrini, Sci. Rep. 3, 1439 (2013).

\bibitem{Syn7}
A. Mari, A. Farace, N. Didier, V. Giovannetti, and R. Fazio,
Phys. Rev. Lett. {\bf 111}, 103605 (2013).

\bibitem{MS99}  A. Hampton and  D. H. Zanette,
Phys. Rev. Lett.  {\bf 83}, 2179 (1999).

\bibitem{MS05} U. E. Vincent, A. N. Njan, and O. Akinlade, Mod. Phys. Lett. B {\bf 19},
737 (2005).

\bibitem{MS03}
X. Wang, M. Zhan, C-H. Lai and H. Gang,
Phys. Rev. E. {\bf 67}, 066215 (2003).

\bibitem{Qiu10}
H. B. Qiu, J. Tian and L-B Fu,
Phys. Rev. A {\bf 81}, 043613  (2010).

\bibitem{MS12}
J. R. Zhang, H. Jiang, Y. Yang, W. S. Duan and J. M. Chen,
Phys. Scr. {\bf 86}, 065602 (2012).

\bibitem{Qiu13}
J. Tian, H. B. Qiu, G. F. Wang, Y. Chen and L-B Fu,
Phys. Rev. E {\bf 88}, 032906 (2013).

\bibitem{Leggett01}
A. J. Leggett, Rev. Mod. Phys. {\bf 73}, 307 (2001).

\bibitem{WI94} D. J. Wineland, J. J. Bollinger, W. M. Itano, and
D. J. Heinzen, Phys. Rev. A {\bf 50}, 67 (1994).

\bibitem{esteve08}
J. Esteve, 
C. Gross, A. Weller, S. Giovanazzi, and M. K. Oberthaler, Nature
(London) {\bf 455}, 1216, (2008).

\bibitem{gross10} C. Gross, T. Zibold, E. Nicklas, J. Est\`eve, and
M. K. Oberhtaler, Nature (London) {\bf 464}, 1165 (2010).

\bibitem{riedel10} M. F. Riedel, P. B\"ohi, Y. Li, T. W. H\"ansch, A. Sinatra, and
P. Treutlein, Nature (London) {\bf 464}, 1170 (2010).

\bibitem{nshii13} C. C. Nshii, M. Vangeleyn, J. P. Cotter, P. F. Griffin, E. A.
Hinds, C. N. Ironside, P. See, A. G. Sinclari, E. Riss, and A. S.
Arnold, Nat. Nanotechnol. {\bf 8}, 321 (2013).

\bibitem{ours10} B. Julia-Diaz, D. Dagnino, M. Lewenstein,
J. Martorell, A. Polls, Phys. Rev. A {\bf  81}, 023615 (2010).

\bibitem{ima97}
A. Imamog\-glu, M. Lewenstein and L. You, Phys. Rev. Lett. {\bf 78}, 2511 (1997).

\bibitem{Milburn97}
G.J. Milburn, J. Corney, E. M. Wright, and D. F. Walls,
Phys. Rev. A {\bf 55}, 4318 (1997).

\bibitem{JAME06} M. J\"a\"askel\"ainen, and P. Meystre, Phys. Rev. A {\bf 71}, 043603 (2005).

\bibitem{baym} E. J. Mueller, T-L. Ho,
M. Ueda, and G. Baym, Phys. Rev. A {\bf 74}, 033612 (2006).

\bibitem{bloch-rmp}
I. Bloch,, J. Dalibard, and W. Zwerger, Rev. Mod. Phys. {\bf 80},
885 (2008).

\bibitem{lewenstein-book} M. Lewenstein, A. Sanpera,
V. Ahufinger, \textit{Ultracold Atoms in Optical Lattices:
Simulating Quantum Many-Body Systems} (Oxford University Press,
2013).

\bibitem{Ashhab02}  S. Ashhab  and C. Lobo,
Phys. Rev. A {\bf 66}, 013609 (2002).

\bibitem{diaz09}
B. Juli\'a-D\'\i az, M. Guilleumas, M. Lewenstein, A. Polls, A.
Sanpera, Phys. Rev. A {\bf 80}, 023616 (2009).

\bibitem{satija09}
I. I. Satija, R. Balakrishnan, P. Naudus, J. Heward, M. Edwards, C.
W. Clark, Phys. Rev. A {\bf 79}, 033616 (2009).

\bibitem{mazzarella10}
G. Mazzarella, M. Moratti, L. Salasnich, F. Toigo, J. Phys. B: At.,
Mol. Opt. Phys. {\bf 43}, 065303 (2010).

\bibitem{naddeo10}
A Naddeo, R Citro, J. Phys. B: At. Mol. Opt. Phys {\bf 43}, 135302
(2010).

\bibitem{mmm11}
M. Mele-Messeguer, B. Julia-Diaz, M. Guilleumas, A. Polls, A.
Sanpera, New J. Phys.  {\bf 13}, 033012 (2011).

\bibitem{weiss}
N Teichmann, C Weiss, EPL {\bf 78}, 10009 (2007).

\bibitem{Fesh}
S. B. Papp and C. E. Wieman, Phys. Rev. Lett. {\bf 97}, 180404 (2006);
S. B. Papp, J. M. Pino, and C. E. Wieman, Phys. Rev. Lett. {\bf 101}, 040402
(2008); G. Thalhammer, G. Barontini, L. De Sarlo, J. Catani, F.
Minardi, and M. Inguscio, Phys. Rev. Lett. {\bf 100}, 210402 (2008);
D. J. McCarron, H. W. Cho, D. L. Jenkin, M. P. K\"oppinger,
and S. L. Cornish, Phys. Rev. A {\bf 84}, 011603(R) (2011).

\bibitem{conf} M. Olshanii, Phys. Rev. Lett. {\bf 81}, 938 (1998).

\bibitem{Albiez05} M. Albiez,  R. Gati, J. F\"olling, S. Hunsmann , M.  Cristiani and
 M. K. Oberthaler, Phys. Rev. Lett. {\bf 95}, 010402 (2005).

\bibitem{berrada13}
T. Berrada, S. van Frank, R. B\"ucker, T. Schumm, J.-F. Schaff,
J Schmiedmayer, Nat. Commun. {\bf 4}, 2077 (2013).

\bibitem{Zibold10}
T. Zibold, E. Nicklas, C. Gross, and M. K. Oberthaler,
Phys. Rev. Lett. {\bf 105}, 204101 (2010).

\bibitem{luc11}
B. L\"{u}cke, M. Scherer, J. Kruse, L. Pezze, F. Deuretzbacher, P.
Hyllus, O. Topic, J. Peise, W. Ertmer, J. Arlt, L. Santos, A.
Smerzi, and C. Klempt, Science {\bf 334}, 773 (2011).

\bibitem{car13}
I. Carusotto and C. Ciuti, Rev. Mod. Phys. {\bf 85}
299 (2013).

\bibitem{gal13} M. Abbarchi, A. Amo, V. G. Sala, D. D. Solnyshkov,
H. Flayac, L. Ferrier, I. Sagnes, E. Galopin, A. Lemaitre, G. Malpuech, and J.
Bloch, Nat. Phys. {\bf 9}, 275 (2013).

\bibitem{li08} Y. Li, Y. Castin, A. Sinatra, Phys. Rev. Lett. {\bf 100}, 210401 (2008).

\end{thebibliography}
\end{document}